\documentclass{webofc}
\usepackage[varg]{txfonts}   
\woctitle{Physics at the magnetospheric boundary}
\begin{document}
\title{The magnetospheric accretion/ejection process in young stellar objects: open issues
  and perspectives}
%
%

\author{J. Bouvier\inst{1}\fnsep\thanks{\email{Jerome.Bouvier@obs.ujf-grenoble.fr}}
}

\institute{ UJF-Grenoble 1 / CNRS-INSU, Institut de Plan\'etologie et
  d'Astrophysique de Grenoble (IPAG) UMR 5274, Grenoble, F-38041,
  France 
         }

         \abstract{%
           This summary talk aims at highlighting some of the
           remaining open issues regarding the physics of the
           magnetospheric accretion/ejection process in young stellar
           objects (YSOs). It lists a number of questions related to YSOs
           magnetic fields and accretion regimes, the structure and
           properties of accretion shocks, the origin of multiple
           outflow components, and the impact of the star-disk
           magnetic interaction onto early angular momentum evolution.
         }
\maketitle
\section{Introduction}

The presentations at this conference have clearly illustrated the new
insights theory, numerical simulations, and observations have brought
to our understanding of the magnetic star-disk interaction process in
young stellar objects (YSOs). I shall rely on these presentations to
outline in this summary talk some of the remaining issues to be
addressed in the forthcoming years regarding the origin and properties
of YSOs magnetic fields, the eventuality of different accretion
regimes, the physics of accretion shocks, the coexistence of several
outflow components, and their impact on angular momentum evolution.

\section{YSOs magnetic fields}

While significant progress has been made on the derivation of magnetic
field strength and topology in YSOs ( cf. contributions to this volume
by S. Gregory, C. Johns-Krull, and S. Hubrig), many aspects remain to
be addressed, including the following ones:
\begin{itemize}
\item What kind of dynamos operates in fully convective and partly radiative young stars? 
\item How does disk accretion impact on the large scale magnetic field topology?
\item What is the short term evolution of magnetic fields (e.g.,
  magnetic cycles, polarity reversals, etc.)? How does it impact on
  accretion flows?
\item How to relate magnetic field properties to stellar parameters
  (mass, age, rotation, Macc, etc.)
\item What is the relative importance of large scale vs. compact magnetic field components?
\item Do accreting and non-accreting young stars share the same magnetic field properties?  
\end{itemize}

On the observational side, large-scale campaigns aiming at measuring
the magnetic field strength and topology of YSOs over a wide mass
range and covering various evolutionary stages, from embedded
protostars to Class III sources, are required to settle these
issues. On the theoretical side, numerical simulations of fully
convective dynamos in rapidly rotating stars, possibly integrating the
interaction with a circumstellar disk as an outer boundary condition,
would provide clues to the magnetic field generation in young stars.

\section{Accretion regimes}

The last 20 years have seen a shift in paradigm from boundary layer
accretion to magnetospheric accretion. Recent observational and
numerical studies, however, tend to suggest that different accretion
regimes may operate in young stars (cf. contributions to this volume
by A.-M. Cody, S. Alencar, N. Fonseca, R. Kurosawa, R. Lovelace, and
M. Romanova). This opens the intriguing possibility that a young star
may experience different accretion phases during its early pre-main
sequence evolution, let alone the rare but intense FU Ori events. In
turn, this new perspective rises a number of issues:
\begin{itemize} 
\item Through how many different ways can accretion occur onto the
  stellar surface: stable, unstable, propeller, quiescent?
\item How to relate the different accretion regimes to the system’s
  properties (accretion rate, magnetic field strength, topology, and obliquity, mass,
  rotation rate, etc.)?
\item What is the observational evidence for different accretion
  regimes? Can different accretion regimes coexist in the same system? 
\item Is accretion steady-state or episodic? On which timescales?
\end{itemize}

Long-term spectrophotometric monitoring of ensembles of YSOs is
essential to better understand the different accretion modes in these
systems. Also, identifying a few prototypical systems exhibiting (or
alternating between) the various accretion regimes would be of great
interest. Obviously, the continuing development of 3D numerical
simulations of the magnetic star-disk interaction will provide strong
insight into these issues, as they explore different mass accretion
rates and magnetic topologies, and include evermore refined physics.

\section{Accretion shocks}

Renewed accretion shock models have been flourishing over the last
years, mostly through the development of 2D numerical simulations of
accretion funnel flows reaching the stellar surface at free fall
velocity (cf. contributions to this volume by L. de S{\`a} and
T. Matsakos). In parallel, X-ray observations of YSOs have recently
provided a new and powerful diagnostics of the physics of accretion
shocks (cf. R. Bonito's contribution to this volume). Current
developments will help addressing the following issues: 
\begin{itemize}
\item What is the structure, stability, and evolution of accretion
  shocks at the surface of YSOs?
\item What observational diagnostics are best suited to investigate
  their physics? What can we learn from X-ray spectra and UV line
  profiles? 
\item What is the relative importance of compact vs. large-scale
  magnetic components in governing the accretion shock structure?
\item Do different accretion regimes translate into different shock
  properties? 
\item Where does the energy dissipated in the accretion shock go
  (stellar interior, outflows, radiation, etc.)? How does it impact
  early stellar evolution and disk chemistry/evolution? 
\end{itemize}

The on-going effort to combine MHD numerical simulations with
radiative transfer computations is sorely needed to be able to provide
quantitative predictions of observable diagnostics, such as continuum
excess flux, emission line flux and profiles, as well as their
expected temporal variability. These predictions will prove extremely
useful to design observations that will efficiently explore the
physics of accretion shocks.

\section{Outflows}

Outflows are intimately related to the accretion process in YSOs. The
tight relationship observed between the strength of accretion and
outflow diagnostics clearly indicates that accretion power drives the
ejection process(es). Beyond this general statement, many outflow
models have been proposed (cf. contributions to this volume by P. Lii,
M. Romanova, C. Zanni, J. Staff, and B. Stelzer) and major issues
remain indeed regarding the identification of the different outflow
components and how each of them relates to the accretion process:
\begin{itemize}
\item How many outflow components are present in an accreting YSO
  system? Where do they originate from? 
\item How can we distinguish (observationally) between accretion
  driven stellar winds, interface outflows (e.g. magnetospheric
  ejections, X-winds), disk winds? 
\item How is each component powered (by accretion)? How does each
  component contribute to the angular momentum budget of the system? 
\item Are outflows episodic or steady-state? On which timescales?   
\end{itemize}

YSOs offer a number of outflow diagnostics (e.g. blueshifted
absorptions in emission line profiles, forbidden emission lines,
ionized jets, etc.) and the radiative transfer post-processing of MHD
outflow simulations starts to yield specific predictions for these
diagnostics. The exploration of the time domain appears today as one
of the most promising directions to improve our understanding of the
accretion/ejection process in YSOs. By monitoring simultaneously
accretion and ejection diagnostics, one may hope to directly relate
outflow events to accretion bursts, and link their characteristic
variation timescales to spatial scales, thus providing clues to the
regions where they originate.

\section{Angular momentum evolution}

One of the greatest challenges in star formation and early stellar
evolution is to understand the low angular momentum content of
newly-born stars. The observational evidence that accreting stars have
on average lower spin rates that non accreting ones suggests that the
magnetic star-disk interaction process acts to brake the central
object (see R. Pudritz's and F. Gallet's contributions to this
volume). Yet, the physical mechanisms(s) by which enough angular
momentum can be removed from the young star in order to prevent it
from spinning up as it accretes material from its circumstellar disk
is far from being understood. Among the most pressing issues, one may
list:
\begin{itemize}
\item What is the net (time averaged) torque exerted on the central
  star by the magnetically-driven accretion/ejection process? Can it
  be negative for a few million years, as observations suggest? What
  are the dominant spin-down processes?
\item Is the spin evolution of pre-main sequence stars monotonic, or
  are there long-term spin-up/spin-down episodes? 
\item How reliable is the evidence for an accretion-rotation
  connection among young stars? How does it depend on stellar
  parameters (mass, accretion rate, magnetic fields, etc.)?
\item What does the instantaneous rotation rate of a young star tell
  us about its accretion history? What is the impact of different
  accretion regimes on the stellar spin? 
\end{itemize}
Long-term 3D MHD simulations of the star-disk interaction are needed
to predict the angular momentum exchange associated to the
accretion/ejection process and the fate of the stellar
spin. Observationally, a clear understanding of the accretion-rotation
connection awaits more precise measurements of mass accretion rates
and magnetic field properties, as well as the determination of the
gaseous disk's inner radius relative to the corotation radius, either
directly through long baseline interferometry or indirectly through
the analysis of spectral features formed at the disk inner edge.

\section{Conclusion}

The magnetic star-disk interaction process is at the heart of YSOs
physics. Indeed, it accounts for most of their observed properties.
Understanding the details of this complex, magnetically-mediated
interplay between the central star and its surrounding disk and
outflows is therefore one of the ultimate goals in star formation
research. The time is ripe for a detailed confrontation between
in-depth modeling (e.g., time-dependent 3D MHD simulations coupled to
radiative transfer models) and observations (e.g., intense
spectrophotometric monitoring of accretion and ejection diagnostics of
selected YSOs). This offers a promising direction to unveil the
structure and physics of the inner 0.1 AU region of young stellar
systems. Within a few years timescale, the magnetic properties of YSOs
will eventually be fully documented, thanks to a combination of
observational techniques including high-resolution spectroscopy and
spectropolarimetry in the optical and the infrared, while the inner
disk structure will be further constrained from long baseline
spectro-interferometry. The parallel advances of observations and
models thus offer the best promise for rapid progress in our
understanding of the magnetic star-disk interaction process in young
stellar systems.

\begin{acknowledgement}
  It is a pleasure to thank the organisers for a fruitful and
  enjoyable conference. This contribution was supported by grant ANR
  2011 Blanc SIMI5-6 020 01 ``Toupies: Towards understanding the spin
  evolution of stars" (\url{http://ipag.osug.fr/Anr_Toupies/}).
\end{acknowledgement}

%
%
%
%
%

\end{document}